\documentclass[a4paper]{jpconf}
\usepackage{graphicx}
\begin{document}
\title{The CRESST II Dark Matter Search}

\author{Leo Stodolsky$^1$, G Angloher$^1$, M Bauer$^3$, I
Bavykina$^1$, A Bento$^{1,5}$, C Bucci$^4$, C Ciemniak$^2$, G
Deuter$^3$, F v Feilitzsch$^2$, D Hauff$^1$,  P Huff$^1$, C
Isaila$^2$, J Jochum$^3$, M Kiefer$^1$,
 M Kimmerle$^3$,  J C Lanfranchi$^2$, S Pfister$^2$,
F Petricca$^1$, W Potzel$^2$, F Pr\"obst$^1$, F Reindl$^1$, S
Roth$^2$, K Rottler$^3$, C Sailer$^3$, K Sch\"affner$^1$, J
Schmaler$^1$, S Scholl$^3$, W Seidel$^1$, M v Sivers$^2$,  C
Strandhagen$^3$, R Strauss$^2$, A Tanzke$^1$, I Usherov$^3$, S
Wawoczny$^2$, M Willers$^2$ and A Z\"oller$^2$
}

\address{$^1$ Max-Planck-Institut f\"ur Physik, F\"ohringer Ring 6,
D-80805 M\"unchen, Germany}
\address{{$^2$} Physik-Department E15, Technische Universit\"at
M\"unchen, D-85747 Garching, Germany}
\address{$^3$ Eberhard-Karls-Universit\"at T\"ubingen, D-72076
T\"ubingen, Germany}
\address{$^4$ INFN, Laboratori Nazionali del Gran Sasso, I-67010
Assergi, Italy}
\address{$^5$ CI, Physics Department, University of Coimbra, P-3004
516 Coimbra, Portugal }

\ead{les@mpp.mpg.de}

\begin{abstract}
Direct Dark Matter detection with cryodetectors is briefly
discussed, with particular mention of the possibility of the
identification of the recoil nucleus.
Preliminary results from the CREEST II Dark Matter search, with 730
kg-days of data, are presented.  Major backgrounds and methods of
identifying and dealing with them are indicated.
\end{abstract}

\section{Introduction}

CRESST II is a cryogenic Dark Matter search operating in the Gran
Sasso laboratory and is a collaboration between the
 Max-Planck-Institute, the Technical University of Munich, the
University of T\"ubingen, and the Laboratori Nazionale del Gran
Sasso.

CRESST is distinguished by the presence of two cryogenic readout
channels. One is for heat or phonons. This provides a very good
measurment of the total energy of an event. The other channel is
for the scintillation
light produced in the target material, which is the scintillating
crystal $CaWO_4$. This light signal is used to greatly reduce the
electron-photon backgrounds. For the nuclear recoils it also 
provides some information on which nucleus is recoiling, which can
play an important role in the analysis.

\section{Why Cryodetectors?}

The proposal to look for direct detection of dark matter was
stimulated by the suggestion of using cryodetectors for weak
processes \cite{gw} and it might be helpful to recall the
motivation for
using
cryodetectors. We are looking for `WIMP'-induced nuclear recoils.
The big problem in detecting this process is the small energy
expected for the recoil.
\begin{figure}[h]
\begin{minipage}{16pc}
\includegraphics[width=16pc]{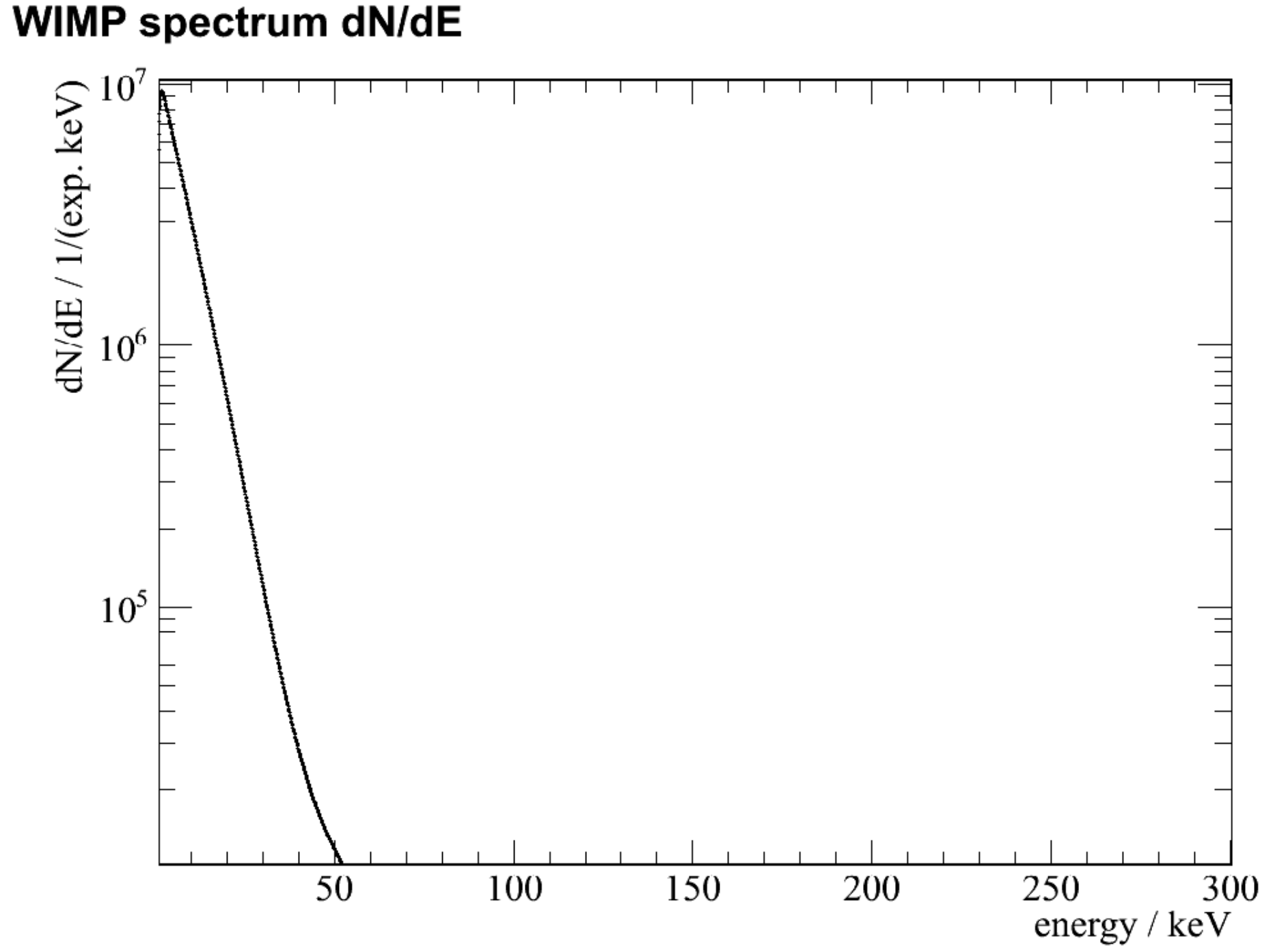}
\caption{Recoil energy spectrum expected in CRESST ($CaWO_4$) for
a M=50 GeV WIMP.}
\end{minipage}\hspace{2pc}%
\begin{minipage}{16pc}
\includegraphics[width=16pc]{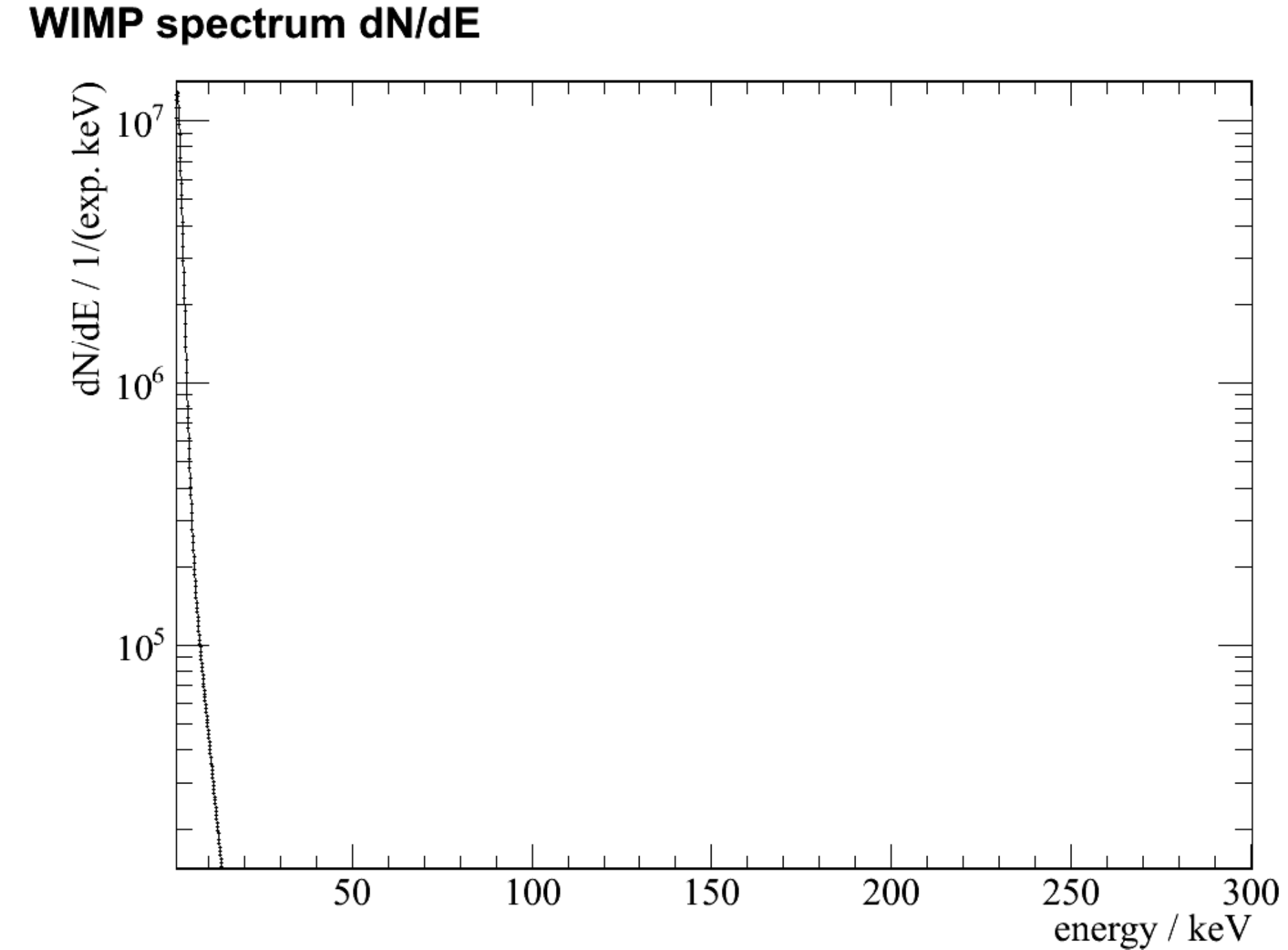}
\caption{Recoil energy spectrum expected in CRESST ($CaWO_4$) for
a M=10 GeV WIMP.}
\end{minipage}
\end{figure}

As one sees from the example illustrated in Fig 1 with a 50 GeV
WIMP, the recoil energy spectrum drops radically with energy.
And the situation is even more dramatic with light WIMPs,
 as we see in Fig 2 with
a 10 GeV WIMP.

It is evident that an optimal dark matter detector should be able
to see and resolve  small ($\leq keV$) energies. However, most
classical particle detection methods are barely able to get down to
this
range, and when employed for dark matter detection are operating on
the ragged edge of their capabilities. For this reason many
detector projects, not employing low temperatures, are effectively
operating way out on the tail of the distributions, or in a
regime where a special understanding of the detector is necessary.
This difficulty, as we see from the figures,
 is especially present for light WIMPs.

\subsection{Energy Threshold and Resolution}
On the other hand cryodetectors are well adapted to this problem.
The reason  may be understood in terms of energy scales
\cite{physto}. In conventional
detectors, whether using liquids, gases, or solid state devices,
via
ionization
or scintillation, the detection process starts by activating or
 ejecting an electron in some way. As is familiar from atomic or
solid state physics, the energy unit for such
processes is 
the electron volt. On the other hand with
cryo-devices we are
dealing with much smaller energies. For superconductivity for
example,  the
typical energy unit is the energy to
break a 
Cooper pair, which is $\sim 1 ^o\,K\sim 10^{-4} eV$ for classical
superconductors and can be even less, as for
the tungsten thermometers in CRESST. 
Thus one has considerably more and finer excitations for our few
keV, implying a much
higher accuracy in the final measurement.

Fig 3 shows the principle of a CRESST detector. A superconducting
film thermometer, held at the superconducting-normal transition,
has a resistance which is sensitive to very small temperature
changes $\Delta T$. These changes are  read out by Squid
electronics as in Fig 4. 

\begin{figure}[h]
\begin{minipage}{16pc}
\includegraphics[width=16pc]{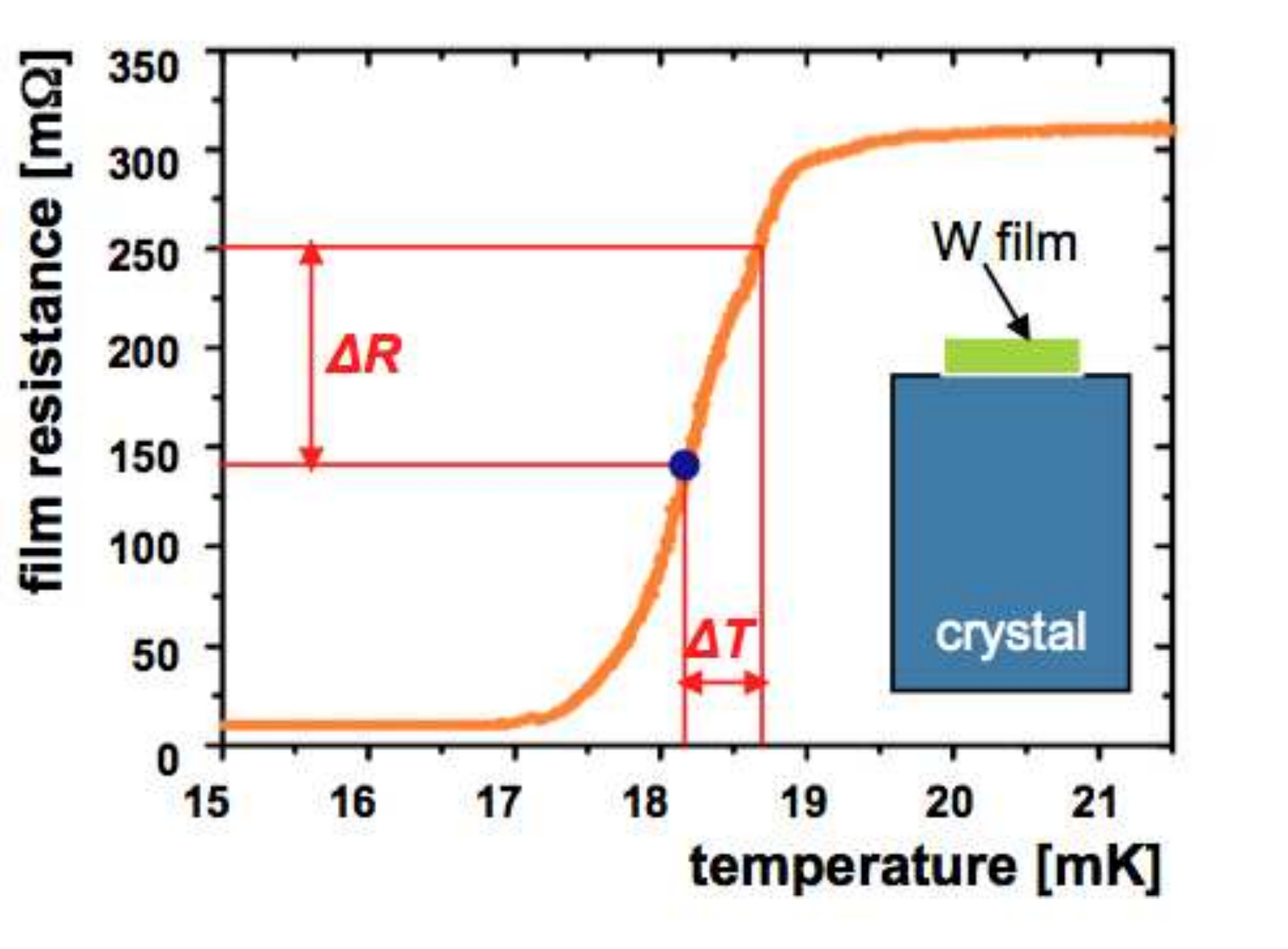}
\caption{ Principle of a CRESST detector with a superconducting
thermometer deposited on a crystal  ($CaWO_4$).}
\end{minipage}\hspace{2pc}%
\begin{minipage}{16pc}
\includegraphics[width=16pc]{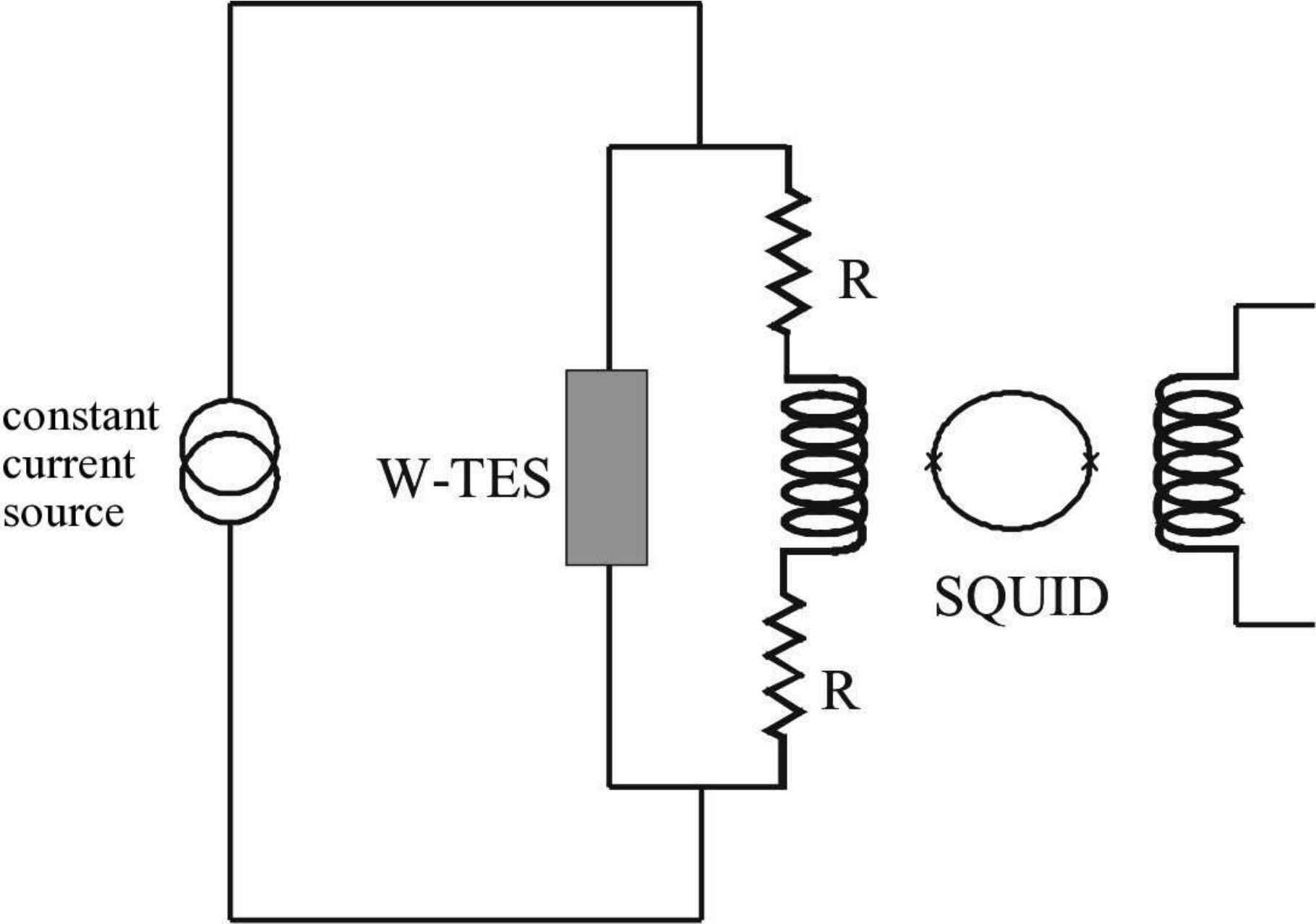}
\caption{Readout circuit for the  superconducting thermometer
(TES).}
\end{minipage}
\end{figure}

The beautiful energy resolution of these detectors is illustrated
\cite{Angloher:2008jj}
 in Fig 5. One notes, around 46 KeV, the  onset of a
$\gamma-\beta$ feature due to the presence of a small $^{210}Pb$
impurity. This onset is very sharp, showing the fine energy
resolution. The other features of the spectrum are also understood.
\begin{figure}[h]
\includegraphics[width=18pc]{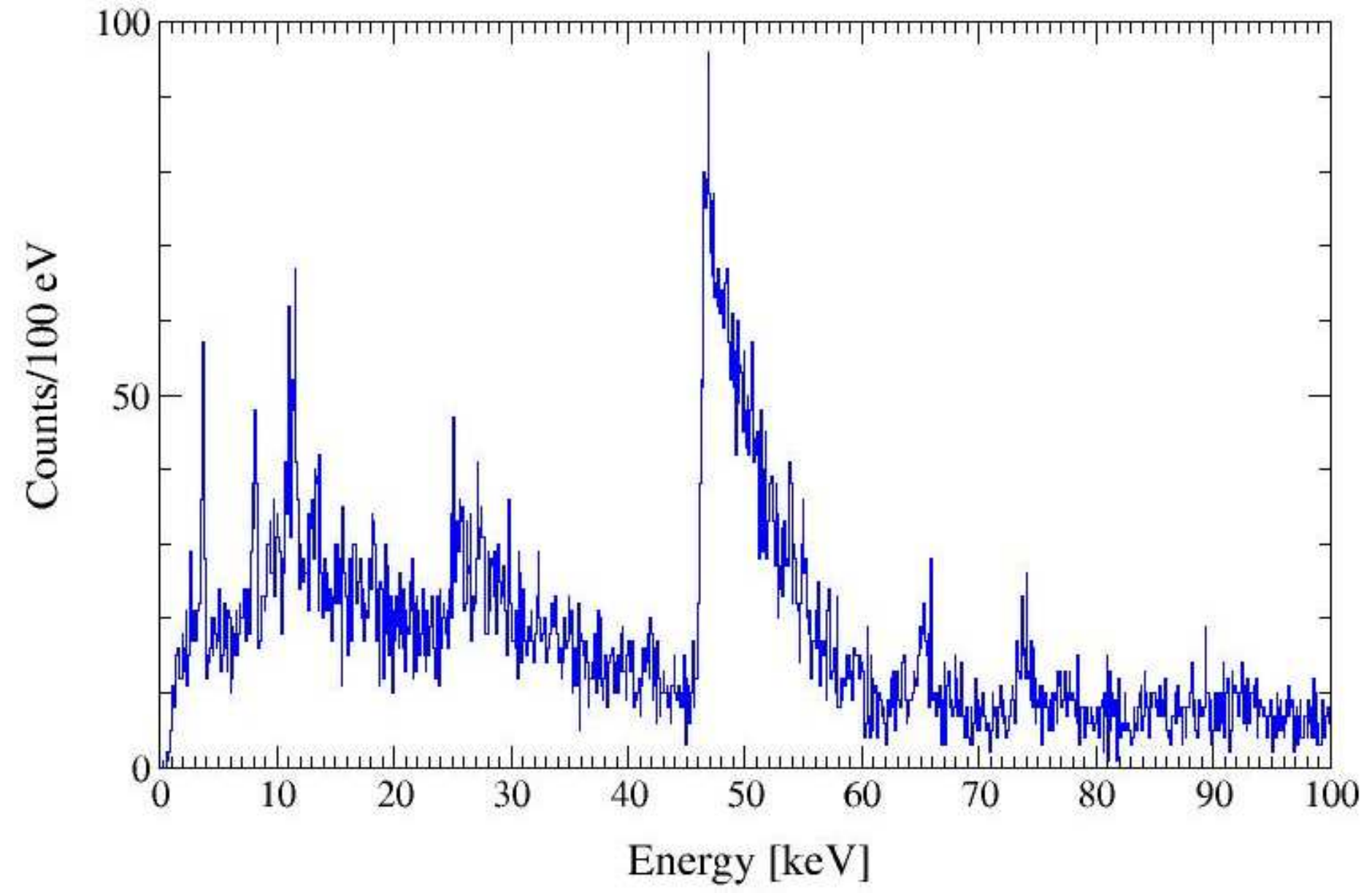}\hspace{2pc}%
\begin{minipage}[b]{14pc}
\caption{ Spectrum from a CRESST detector illustrating the very
good energy resolution, as exhibited by the  features due to
small impurities.}
\end{minipage}
\end{figure}

There are two points to be noted here. One is the low  energy {\it
threshold}, which allows one to get down to  the main part of the
expected recoil energy spectrum. Secondly, there is the very good
energy {\it resolution}. It will be appreciated from Figs 1 and 2
that, due to the very rapid variation with energy, even a small
error in an energy determination can lead to
a very big error in the rate to be expected. Thus even in the
absence of a positive signal, where we are just setting upper
limits, a good understanding of the energy
resolution is necessary to  quantitatively determine what WIMP
parameters are being
excluded. This is especially true for light WIMPs. 

\subsection{Multi-element targets}
 There is another advantage to cryodetectors. The cryo-technique is
well adapted to studying different nuclear
targets at the same time; that is, in parallel in the same setup.
For example,
in CRESST we have  oxygen, calcium, and tungsten nuclei
present simultaneously and their recoils are read out at the same
time  by the same system. In addition, the
superconducting thermometer, or perhaps another cryo-sensor, can
be applied onto
various materials. These features are in strong contrast to
detectors using
only one element, such as a noble gas or liquid, which are designed
around the properties of that one element. While such
detectors  have certain advantages, the multi-target aspect
 is definitely missing.

This ability to compare different nuclei gives valuable extra
information and could play an important role in verifying a
positive signal. Given
the suggestion of a positive signal, there will always be the
suspicion that it is due to some unsuspected background, and this
fear may
not be unfounded, given the very low rates involved and
the rather unspecific nature of the WIMP signal. It would therefore
be very helpful  to have some features of the data which are
characteristic of the sought-for WIMP signal.

 A good, and one of the few, possibilities for something
characteristic would be the comparison
of the properties of the signal on different nuclei. This should
vary in a definite way as we look at different  nuclei, something
not in general true for most backgrounds.

\subsubsection{Recoil Spectrum}
One of the simplest features concerns the shape of the recoil
energy spectrum. Since the incoming WIMP flux is evidently the same
for all
our target nuclei, the shape of the recoil  energy
spectrum should vary in a well defined way from nucleus to 
nucleus. This shape is just given by the mass and velocity spectrum
of the incoming WIMPs, which is the same for all nuclei, and the
mass and form factor of the nucleus. In addition to the mass, the
form factor is the only quantity varying with the nucleus. It is
however, a well known
quantity, and for the small recoils where most of the rate is
concentrated, depends on only one parameter, the radius of the
nucleus.

Thus  the shape of the recoil spectra on different nuclei should be
in agreement with each other, once a mass for the WIMP is assumed,
and 
observation of the correct
behavior would go a long way towards making a WIMP signal
convincing.

We stress the shape of the recoil spectrum and not its absolute
level or total rate. The absolute rate 
 can of course vary from nucleus to nucleus, depending on the
quantum numbers of the WIMP and the number of neutrons and protons
in the nucleus, and whether we have coherent or spin-dependent
interactions.
 In fact such absolute variations could be used to disentangle the
composition of the WIMP\cite{Gabutti:1996qd}.
But the shape is governed by the simple factors just mentioned.

\subsection{Annual variation}
 Similarly the annual variation effect\cite{ann}, coming from  the
different velocities the earth has with respect to the galactic
halo in summer and winter, is due to a simple variation of the
incoming flux and should
be essentially the same for all nuclei. Thus here also the
observation of the
same percentage variation on the different nuclei of a target
material would help greatly in establishing the effect.

\subsection{Fast neutrons}
 A particular difficulty would be a fast neutron background. Fast
neutrons would induce a recoil spectrum like that from the nuclear
form factor. And if an annual variation 
effect is due to  fast neutrons  created by cosmic muons, 
these would also show the same relative variation for different
nuclei.
However--and here we again see the advantage of being able to
distinguish which nucleus is recoiling--fast neutrons undergo
diffraction scattering and diffraction scattering  shows a
characteristic variation from one nucleus to
another, different from that expected for WIMPs. This is discussed
in ref\,\cite{hen} and in table 1 we
illustrate the point by showing  for different
nuclei the rate, relative to that on oxygen, for
coherently scattering WIMPs and for diffractively scattering
neutrons. One notes  different patterns in the variation with 
nucleus. This could be used to distinguish WIMPs from neutrons.

\begin{table}\label{tab1}
\begin{minipage}{12pc}
\caption{ The differential scattering rate on various nuclei in
the diffractive `black disc'
limit, 
at $E_{r}=$20\,keV and  $E_{r}=$30\,keV. 
The same
rate for a   coherently scattering WIMP at $E_{r}=20\,keV$ 
for masses 
 10 and 50 GeV is also shown. One notes  different patterns of A
behavior for neutrons and WIMPs.  All
values are per nucleus and normalized to that for oxygen. From
ref\,\cite{hen}. } 
\end{minipage} \hspace{1pc}%
\begin{minipage}{14pc}
\begin{tabular}{|l|l|l|l|l|l|}
\hline
$Element$&$A$&{neutron}&neutron&
WIMP&WIMP\\
&        &   {$E_{r}=$20\,keV}&$E_{r}=$30\,keV&M=10 GeV& M=50 GeV\\
\hline
\hline
O&16&1&1&1&1\\
\hline
F&19&1.5&1.5&1.3&1.8\\
\hline
Na&23&2.2&2.2&1.6&3.3\\
\hline
Si&28&3.4&3.3&1.8&6.7\\
\hline
Ar&40&7.0&6.4&1.1&19\\
\hline
Ca&40&7.0 &6.4&1.1&19\\
\hline
Ge&74&19&13&$\sim\,$0&93\\
\hline
I&127&20&5.1&$\sim\,$0&200\\
\hline
Xe&132&18 &3.9&$\sim\,$0&240\\
\hline
W&184&2.6 &1.6&$\sim\,$0&230\\
\hline
\end{tabular}
\end{minipage}
\end{table}

\section{The Light Channel}
In CRESST, information about the nature of the recoil is obtained
via  the scintillation light, the ``light channel". Fig 6 shows a
module with the target crystal and light detector, all surrounded
by
a reflecting and scintillating foil.

\begin{figure}[h]
\includegraphics[width=12pc]{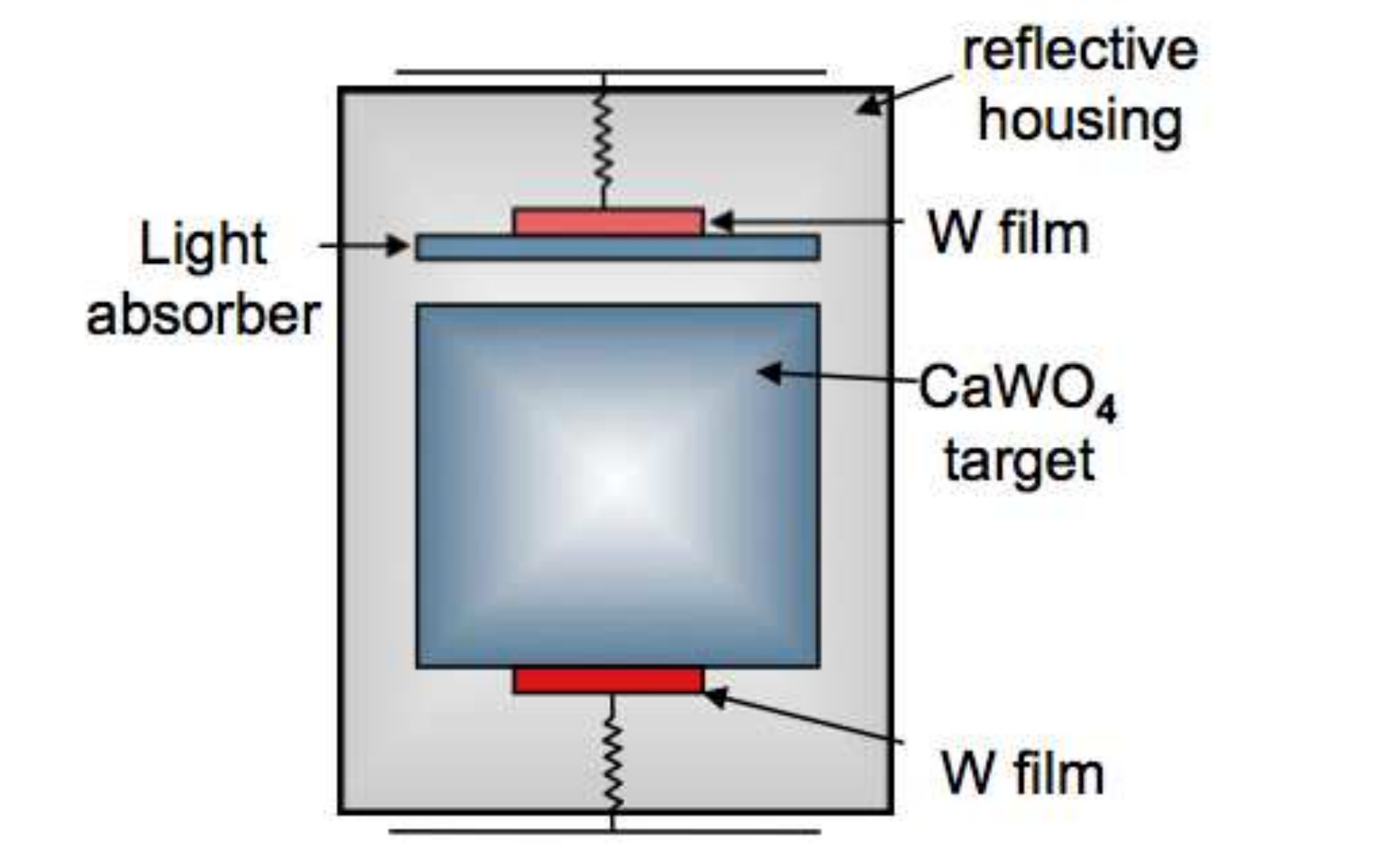}\hspace{2pc}%
\begin{minipage}[b]{14pc}\caption{A module with main detector and
light detector, surrounded by a reflecting and scintillating
housing.}
\end{minipage}
\end{figure}

 The distinction
as to the type of recoil particle  is based on the fact that the
ratio (light output/energy of the event) is high for fast, light,
particles and  low for slow, heavy, particles. This distinction is
quantified in term of the ``quenching factor'', QF, which is
defined as the ratio of the light output from a given nucleus to
that for an electron of the same energy. The QF has been
extensively
studied by the CRESST collaboration, with the values
$QF_{\alpha}\approx  0.22,\;QF_{O}\approx  0.10,\;
QF_{Ca}\approx  0.064,\; QF_{W}\approx  0.040,$ in $CaWO_4$.
There are in general large fluctuations around these average
values, as one sees in the plots below. Events from a run  are
plotted in the (energy, light yield) plane, where `light yield'
is defined as the light output relative to that for  122\,keV
photons from a  $^{57}Co$ source used for calibration.

 First of all, the light channel is used to separate the very large
electron-photon background from nuclear recoils. This seems to work
quite well, as we illustrate by comparing a run with a neutron test
source Fig 7, and a run without the test source Fig 8, from
ref\,\cite{Angloher:2008jj}. One sees that in the latter the
neutron-induced nuclear recoils have disappeared. WIMP candidates
will thus appear on such plots as events with low energy and low
light yield.

\begin{figure}[h]
\begin{minipage}{16pc}
\includegraphics[width=19pc]{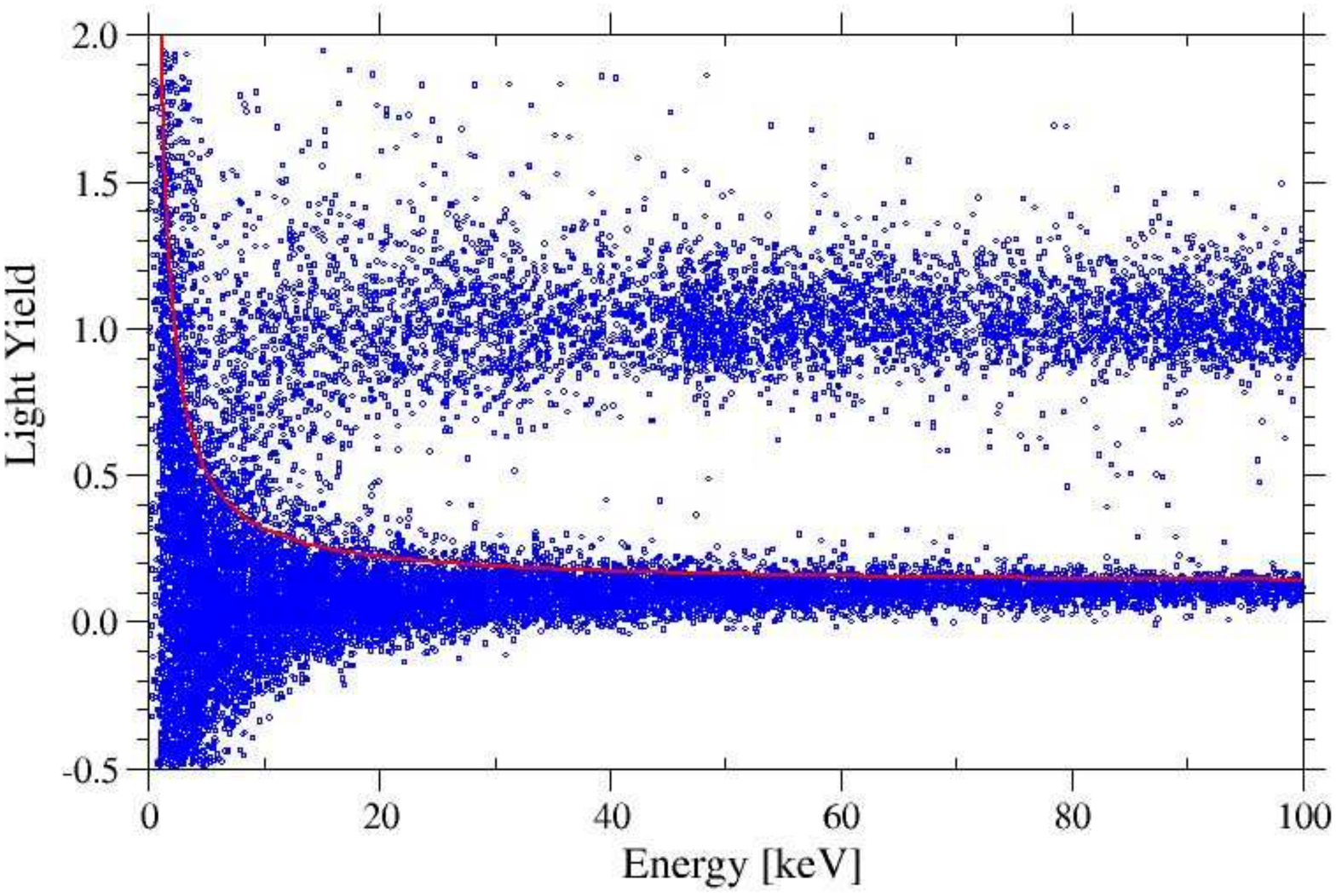}
\caption{ Events in the light yield-energy plane with a neutron
source present. Two bands are seen, one for $e/\gamma$ events, the
other for neutron-induced nuclear recoils.}
\end{minipage}\hspace{2pc}%
\begin{minipage}{19pc}
\includegraphics[width=19pc]{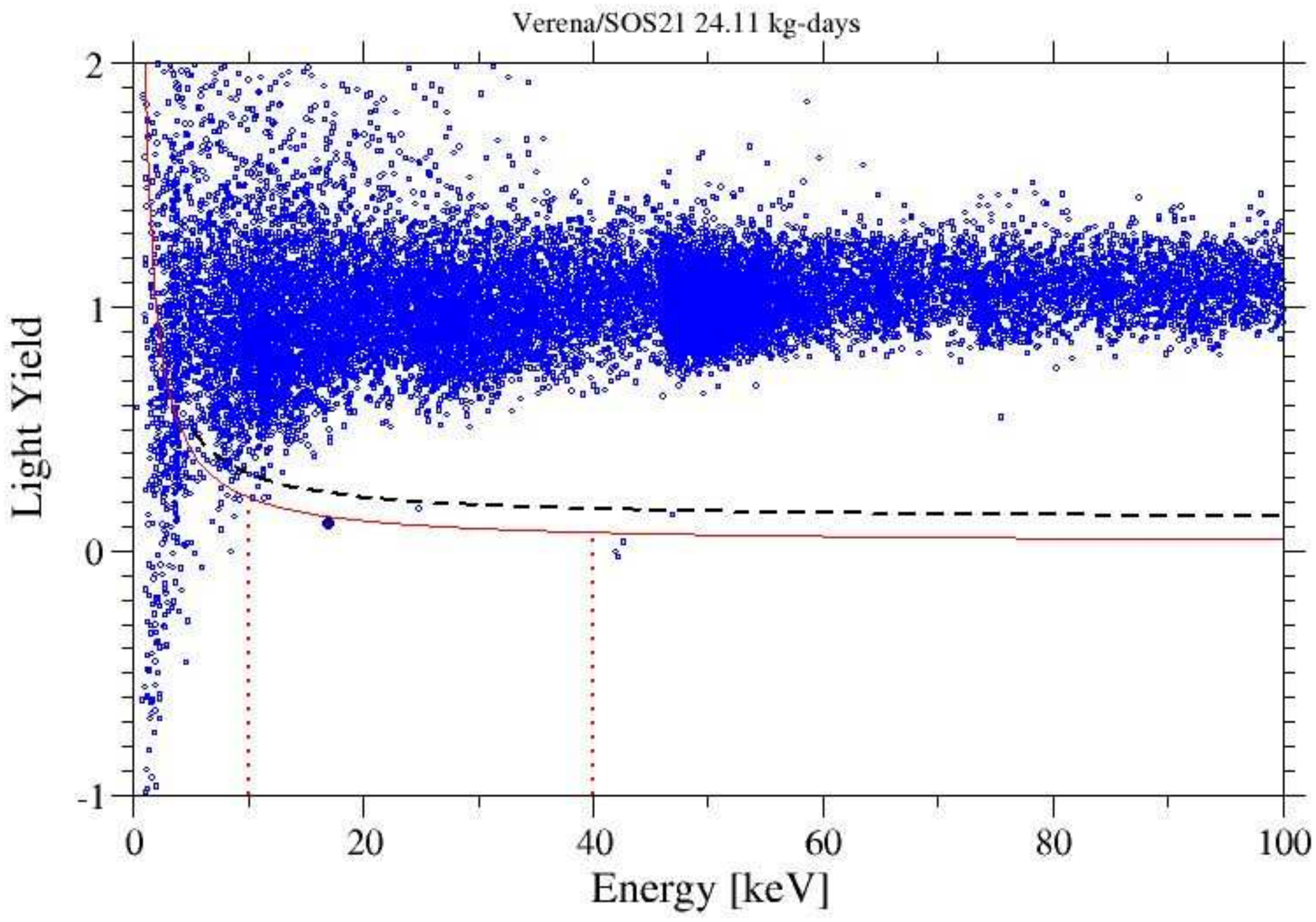}
\caption{The same with the neutron source removed. The lower
 band is now absent. The event marked with a larger dot would be
a WIMP candidate. `Light yield' is defined as the light output
relative to that for a 122\,keV photon in the same detector. }
\end{minipage}
\end{figure}

With the present detectors, the separation of the recoils of the
different nuclei from each other is not as clear as it is for the
$e-\gamma$ band. Since the QF get small, there is little light and
the separation becomes difficult at low energy.

\section{Results from 730 kg-days}
In the data analysis of a dark matter run an acceptance region at
low energy and light yield is defined. This is shown for one module
in Fig 9.
\begin{figure}[h]
\includegraphics[width=22pc]{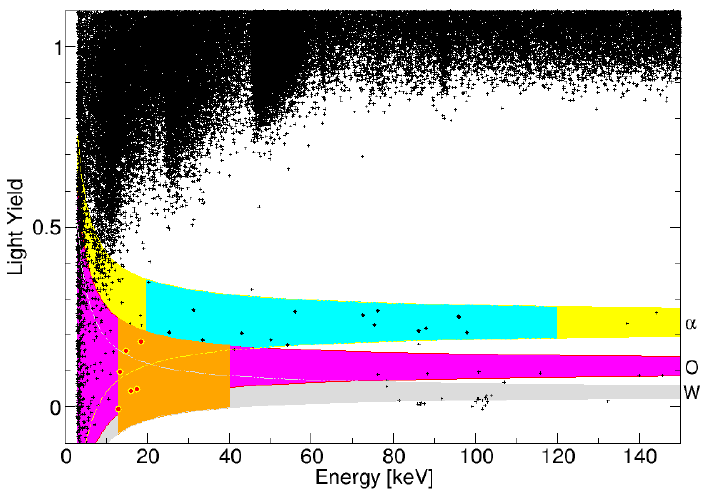}\hspace{2pc}%
\begin{minipage}[b]{14pc}\caption{ Events in the light yield-energy
plane from one of the modules from a long dark matter run, with a
total of 730 kg-days for all 8 modules. The acceptance region for
WIMP
scattering candidates is colored  orange and for this module
contains 6 events (red dots). Also shown are the expected (90\%)
bands for $\alpha$'s, and oxygen and tungsten nuclei. The  calcium
band is not
shown but is included in the analysis.}
\end{minipage}
\end{figure}
The lower boundary in energy is set so that only one
``leakage'' event  from the $e-\gamma$ band is expected; the upper
boundary is set at 40 keV, where according to calculations as in
Fig 1, neglible  WIMP rates are expected. The boundaries in light
yield are chosen to include oxygen, calcium, and tungsten recoils.
In all, from  eight modules, the acceptance regions are found to
contain 67 events.
\section{Background}
 The question then is: can all the events in the acceptance regions
can be explained as
background? Three forms  of  background are evident in Fig 9. First
there is the ``leakage''  from the $e-\gamma$ band, whose expected
value has been used to set the lower boundary of the acceptance
region. Secondly and thirdly there are $\alpha's$ and $Pb$ ions
from the decay $^{210}Po\to ^{206}Pb\,(103\,keV) +
\alpha\,(5.3\,MeV)$. We believe these originate in the clamps
holding the crystals. The $\alpha$'s and $Pb$ ions can lose energy
on their way out of the clamps and so possibly `leak' down into the
acceptance  region. The probability of this occuring can be
estimated by taking the observed events  outside the acceptance
region and extrapolating into the acceptance region. For example
the blue region in Fig 9 is used to determine the $\alpha$
spectrum.

 Finally there is a fourth and perhaps most difficult background,
one not seen on the plot: neutrons. The scattering of neutrons
on nuclei can resemble that expected from WIMPs \cite{hen}. But
there is one important difference. Neutrons, being  strongly
interacting, are expected to multiple scatter, while WIMPs, of
course, should not.
The neutron possibility was examined in two ways. One was with a
neutron test source, as in Fig 7. The other was using the muon
veto. CRESST has a muon veto used to exclude events which are in
coincidence with  an incoming muon. However one may also use some
of these coincidences to see the effects of neutrons induced by
muons. From both of these methods, the neutron source or the muon
coincidences, one  may determine the ratio of multiple hits in
different detector modules to single hits. Given this information
one may use the number of multiple hits found in the dark matter
run to scale up to the number of WIMP-like single hits to be
expected from neutrons. In the total dark matter run there were
three events with multiple hits, which when scaled up, imply
relatively few neutrons. 

The sum of the estimates for the backgrounds does not appear to be
able to explain all the  events. In the analysis we thus also
include a possible WIMP signal, assuming coherent scattering for
the WIMP, as throughout. The results of an elaborate maximum
likelihood analysis is shown in table 2. 
\begin{table}\label{tab2}
\begin{minipage}{14pc}
\caption{Backgrounds and possible WIMP signal resulting from 
maximum likelihood fits. }
\end{minipage}
\begin{minipage}{16pc}
\begin{tabular}{l || c | c }
 \hline
  & M1 & M2 \\ \hline
  ${e/\gamma}$-events & $8.00 \pm 0.05$ & $8.00 \pm
0.05$ \\[1ex]

  ${\alpha}$-events & $11.5\,_{-2.3}^{+2.6}$ &
$11.2\,_{-
2.3}^{+2.5}$\\[1ex]

{  neutron} events & $7.5\, _{-5.5}^{+6.3}$ &
$9.7\,_{-5.1}^{+6.1}$\\[1ex]
 { Pb} recoils & $14.8\,_{-5.2}^{+5.3}$ &
$18.7\,_{-4.7}^{+4.9}$\\[1ex]
\hline
  {signal} events & $29.4\,_{-7.7}^{+8.5}$ &
$24.2\,_{-7.2}^{+8.1}$
\\[1ex] \hline\hline

  $m_\chi$ [GeV] & 25.3 & 11.6\\

  $\sigma_{\small WIMP}$ [pb] & $1.6\cdot10^{-6}$ &
$3.7\cdot10^{-5}$\\
\hline
 \end{tabular}
\end{minipage}
\end{table}

The fitting procedure finds two minima M1 and M2 for the likelihood
function, both of about the same strength (4.7 and  4.2 $\sigma$),
with WIMP masses of 25 GeV and 12 GeV respectively. The WIMP cross
section for M1 can be appreciably smaller than for M2 since the
heavier mass allows the scattering on the tungsten to be above
threshold and the coherence $A^2$ factor has a large effect.

It must be said that maximum likelihood analyses should be
appreciated with care. The analysis takes place within the context
of an assumed model, and the resulting $\sigma$'s characterize how
sharply the parameters used in the model are determined. It is thus
perfectly possible to have a bad model with a good sigma.
For this reason some kind of additional `goodness of fit'
characterization is needed. We have examined  the ``p-value''
\cite{pval} which estimates how probable it is that the model 
parameters found from the maximum likelihood fit  would give rise
to the actually  observed data. The value of p turns out to be the
not very small 0.35, which appears acceptable.

\section{Future Plans}
Further analysis using other procedures is underway, and these may
yield somewhat different results. But it is intriguing that the
obvious backgrounds do not seem to entirely explain the data. To
clarify the situation, a further CRESST run, in which it is hoped
there will be substantially reduced backgrounds,  is in
preparation. Since it is believed that the 
$^{210}Po$ background mentioned above originates in the clamps
holding the crystals, a special effort is devoted to producing
clamps of highly pure material.

 Also the installation of a further layer of polyethlyne in the
inner region around the detectors is being studied in order to
provide additional  shielding against neutrons originating in the
inner parts of the setup.

 Finally, efforts aimed at improvment of the light detection, which
as discussed above, could play an important role in the
identification of the recoil nucleus, continue to be a focus of
detector development.

\end{document}